\documentclass[aps,pra,twocolumn,showpacs,notitlepage,longbibliography,floatfix,superscriptaddress,nofootinbib]{revtex4-1}
\usepackage{amsfonts}
\usepackage{mathtools}
\usepackage{graphicx}
\usepackage{epsfig}
\usepackage{dcolumn}
\usepackage{bm}
\usepackage{amsmath}

\usepackage[normalem]{ulem}
\usepackage[latin1]{inputenc}
\usepackage{ulem}
\usepackage{epstopdf}
\usepackage{subfigure}
\usepackage{color}
\usepackage[dvipsnames]{xcolor}
\usepackage{amsthm}
\usepackage{newlfont}
\usepackage{graphicx}
\usepackage{amssymb}
\usepackage{epstopdf}
\usepackage{appendix}
\usepackage[breaklinks=true]{hyperref}
\usepackage{breakcites}
\usepackage{textcomp}
\usepackage{appendix}
\usepackage{multirow}	
\usepackage{color}
\usepackage{amssymb}
\usepackage{epsfig}
\usepackage{bm}
\usepackage[american]{babel}
\usepackage{braket}
\hypersetup{colorlinks=true,linkcolor=blue,citecolor=blue,filecolor=blue,urlcolor=blue,pdfstartview=FitH}
\begin{document}
\title{\textbf{Controlling Vortex Lattice Structure of Binary Bose-Einstein Condensates via Disorder Induced Vortex Pinning}}
\author{ Dibyendu Kuiri}
\thanks{Present address: AGH University of Krakow, Academic Centre for Materials and Nanotechnology, al. A. Mickiewicza 30, 30-059 Krakow, Poland}
\affiliation{\textit Department of Physics, SP Pune University, Pune 411007, India}
\author {Thudiyangal  Mithun}
\affiliation{Department  of  Atomic and Molecular Physics,  Manipal Academy of Higher education, Manipal 576 104, India}
\affiliation{\textit Department of Mathematics and Statistics, University of Massachusetts, Amherst,
MA 01003-4515, USA} 
\author{Bishwajyoti Dey}
\affiliation{\textit Department of Physics, SP Pune University, Pune 411007, India} 

 \begin{abstract}
We study the vortex pinning effect  on the vortex lattice structure of the rotating two-component Bose-Einstein condensates (BECs)  in the presence of impurities or disorder by numerically solving the time-dependent coupled Gross-Pitaevskii equations. We investigate the transition of the vortex lattice structures by changing conditions such as angular frequency, the strength of the inter-component interaction and pinning potential, and also the lattice constant of the periodic pinning potential. We show that even a single impurity pinning potential can change the unpinned vortex lattice structure from triangular to square or from triangular to a structure which is the overlap of triangular and square.  In the presence of periodic pinning potential or optical lattice, we observe the structural transition from the unpinned vortex lattice  to the pinned vortex lattice structure of the optical lattice. In the presence of random pinning potential or disorder, the vortex lattice melts following a two-step process  by creation of lattice defects, dislocations, and disclinations,  with the increase of rotational frequency, similar to that observed for single component Bose-Einstein condensates. However, for the binary BECs, we show that  additionally the two-step vortex lattice melting also occurs  with increasing strength of the inter-component interaction.   
 \end{abstract}
 
 \maketitle
\pagestyle{myheadings}
\section{Introduction}
Studies of multi-component Bose-Einstein condensates (BECs), either of the same atomic species \cite{Myatt,Hall,Maddaloni,Tojo,Egorov} or of different atomic species \cite{Modugno,papp,Thalhammer,McCarron} have become subject of recent interest. This is because of the fact that the presence of two competing energy scales of intra- and inter-component interaction, the multi-component BEC presents novel and fundamentally different ground state scenarios and vortex lattice structures than that of the single-component BECs.  Binary BECs have been realized in various setups: a single isotope in two different hyperfine states \cite{PhysRev}, two different isotopes of the same alkali metal \cite{PhysRevLett2} or two distinct elements \cite{papp,Lercher}.  
By varying the particle numbers of the components, it is possible to go continuously from regimes of inter-penetrating superfluids to those with separated phases \citep{Ho}. The equilibrium vortex lattice structure of rotating single-component BECs is the well-known Abrikosov vortex lattice or triangular (hexagonal) vortex lattice.
On the other hand, a rich variety of vortex lattice structures occurs in rotating multi-component BECs, such as interlaced square  vortex lattice which has been observed in rotating spinor BECs \citep{square}. It has been shown that by varying the strength of the inter-component interaction for binary BECs the interlocked vortex states undergo a phase transition from triangular to square lattices, then to double-core lattices and finally leading to non-periodic interwoven serpentine vortex sheets \citep{Kasamatsu,Review1,Kasamatsu1,Kasamatsu2}. More recently, it has been shown that in rotating binary dipolar BECs new vortex lattice structure is caused by the long-range interactions \citep{Ghazanfari}. For unequal masses of two species  of binary BECs, the two condensates rotate at different speeds due to the disparity in masses and  show vortex synchronization leading to the formation of  bound pairs and the locked state of the two vortex lattices \citep{Barnett_2008,Barnett_2010}. For attractive interaction between the two components of the binary BECs, the system exhibits  non-triangular geometry of the vortex lattices, such as square and two-quantum-vortices \citep{pekko}. More recently, it has been shown that for unequal masses of the atoms, exotic vortex lattice configurations which include the honeycomb, Kagome, and herringbone can exist in binary repulsive BECs \citep{Mingarelli}.

In the past few years, the study of the effect of impurity pinning potential on vortex dynamics in BECs has gained importance. 
A few examples are, vortex lattice melting in the presence of random impurities or disorder and its usefulness to study melting problems, in general, \cite{Mithun_2016, Guillamon}, Anderson localization \cite{Billy2008}, superfluid behaviour \citep{superfluid}, turbulent dynamics in BECs induced by stirring mechanism due to time-dependent impurity position \cite{Mithun2021,Bradely} etc.  Rotating BECs with impurities provides a system where it is possible to display, in a controlled way, the interplay between interaction and disorder in the vortex dynamics. This competition is responsible for the vortex lattice melting in BECs which mimics the observed  vortex lattice melting in type-II superconductors \citep{Mithun_2018}. Such melting is fundamentally different from the more conventional thermal melting. In this case, the transition  can be driven by vortex pinning due to point disorder rather than temperature. It was originally proposed in the context of vortex matter in high-temperature superconductors \citep{Kierfeld,Vinokur}. High-temperature superconductors have two order parameters, the s-wave and d-wave order parameters and the superconductivity of the high-temperature  superconductors depend on impurity doping \citep{Dagotto}. The vortex dynamics in high-temperature superconductors are described by two-components Ginzburg-Landau theory in the presence of applied magnetic field \cite{Karmakar}, \citep{Madhuparna,*Madhuparna1,*Madhuparna2}. The melting process of the vortex lattice in low-temperature superconductors has been studied extensively to explore its role in the critical  current of such superconductors. In this context, the study of the impurity-induced vortex pinning and vortex lattice structure in binary BEC is important as it might mimic the vortex lattice dynamics in high-temperature superconductors which have two order parameters. The identical  idea of melting due to random pinning has also been used by Tsiok {\it et al} and recently \cite{Tisok} among others. 
Likewise, the effects of periodic impurities  on vortex dynamics in BECs have generated great interest recently due to their applicability in various fields such as the physics of Josephson junction arrays, fractional quantum Hall effect, etc. These studies are done by loading the BECs on a rotating optical lattice and experiments employing BECs in a rotating optical lattice have led to the observation of some of these physical phenomena \cite{Tung,Williams}. For weak optical lattice potential, the pinning of vortices by the optical lattice has shown rich vortex lattice structures \cite{Kenichi, Pu, Sato, Kato, Morsch}. 
Similar studies on honeycomb optical lattice have shown interesting moving vortex phases which are useful for studying the anomalies in the critical current of type-II superconductors \cite{Reichhardt}.
Very recently, experimental and theoretical studies of ultracold gases on quasi-crystalline optical lattice potential have been reported \cite{Viebahn,Johnstone}. Quasi-crystalline potentials have long-range order but are not periodic. However, there are very few studies related to the vortex lattice structures of the binary BECs in presence of periodic pinning or optical lattice where the presence of the inter-component interaction further enriches the vortex lattice structures \cite{Reijnders,Mink}.

In this paper, we study the effects of impurities or disorder on the equilibrium vortex lattice structures of the rotating binary BECs. The presence of the impurity potential adds another energy scale to the problem besides the other two competing energy scales of intra- and inter-component interactions in binary BECs. Competitions between these energy scales  allow controlling  the equilibrium vortex lattice structures of the rotating binary BECs via disorder-induced vortex pinning. We show that even a single impurity can change the vortex lattice structure from triangular to square and also from triangular to a distorted lattice.  The structure of the pinned vortex lattice depends on the commensurate or incommensurate positions of the impurities w.r.t. the positions of the vortices of the unpinned lattice, i.e. vortex lattice without impurities. Maximum changes in the unpinned vortex lattice structures occur when the impurities are in incommensurate positions. In the presence of periodic impurities or periodic pinning potentials, which can be created by optical lattice, we show that the vortex lattices acquire the structure of the optical lattice due to the pinning of the vortices by the optical lattice potential. 
In the presence of random impurities or disorder, the vortex lattice melts following a two-step melting process by creating an increasing number of lattice defects, dislocations, and disclinations, with increasing rotational frequency and strength of the random pinning potential. Further, it is shown that the vortex lattice melting of binary BEC in the presence of the disorder is also possible with increasing  strength of the inter-component interaction. To characterize the equilibrium structure of the vortex lattices, we calculate the condensate densities of the components of the binary BEC and its corresponding structure factor profiles. To find the lattice defects in the disordered vortex lattices we plot the Delaunay triangulated disordered vortex lattice showing the lattice defects.
\section{Theoretical model of rotating binary BEC and the coupled Gross-Pitaevskii equations for the system}
We begin with the effective 2-dimensional (2D) Gross-Pitaevskii (GP) energy functional $E[\psi_1, \psi_2] = \int \mathcal {E}_{2D}(\textbf{r}) d^2r$ expressed in terms of the binary condensate wavefunctions $\psi_j$ for the $j$-th component ($j=1,2$), where the energy density is given by
\begin{eqnarray}
\mathcal{E}_{2D}(\textbf{r})&&= \sum^2_{j=1} \bigg( {\hbar^2 \over 2m_j}|\nabla \psi_j|^2 + V_j|\psi_j|^2 + {g_{jj} \over 2} |\psi_j|^4\nonumber\\
 && - \Omega \psi_j^\ast L_z \psi_j \bigg)+ g_{12}|\psi_1|^2|\psi_2|^2,
\label{eq:GPE_ener}
\end{eqnarray}
where $\psi_j^\ast$ is the complex conjugate of $\psi_j$. Here, $m_j$ represent the atomic mass of the $j$-th component, $g_{jj}={4\pi\hbar^2a_j \over m_j} $ the intra-component interaction strength, $g_{12}={2\pi\hbar^2a_{12}\over m_{12}}$ the inter-component interaction strength, $m_{12}={m_1m_2 \over m_1+m_2}$, $a_j$ and $a_{12}$ denote the corresponding $s$-wave scattering lengths, $\Omega$ is the rotational frequency, $L_z$ is the angular momentum in the $z$ direction, with normalization condition $\int(|\psi_j|^2dx dy=N_j$, and $N=N_1+N_2$ the total number of particles in the system. The potential $V_j(x,y)$ consists of two parts $ V_j(x,y)=V_{j_{,trap}}(x,y)+V_{j_{,impurity}}(x,y)$, the harmonic trap potential and the impurity potential respectively. The harmonic trap potential has the form $V_{j_{,trap}}(x,y)={1\over2}m_j\omega_{\perp}^2(x^2+y^2)$, where $\omega_{\perp}$ is the radial harmonic frequency. For a single  impurity at position $(x_0,y_0)$, we take the impurity potential as $V_{j_{,impurity}}(x,y) =V_{0j}\exp\left \{{-[(x-x_0)^2+(y-y_0)^2]\over (\sigma/2)^2}\right\}$, where $V_{0j}$ denote the strength of the impurity potential interacting with $j$-th component and $\sigma$ is the width of the potential. For the periodic distribution of  impurities that can be created by the optical lattice, the impurity potential is taken as the optical lattice potential  $V_{impurity}=V_{lattice}(\textbf{r})=\sum_{n_1,n_2}V_{0}\exp\left\{{-[|\textbf{r} - {\textbf{r}_{{n_1},n_{2}}\mid^2}\over (\sigma/2)^2}\right\}$, where ${\textbf{r}_{n_1,n_2}}=n_1{\textbf{a}_1}+n_2{\textbf{a}_2} $ denote the lattice points, $n_1$ and $n_2$ are integers. For the triangular optical lattice, the two lattice unit vectors are given by ${\textbf{a}_1}=a(0,1)$ and ${\textbf{a}_2}=a(\pm1/2,\sqrt 3/2)$ and for the square optical lattice ${\textbf{a}_1}=a(1,0)$ and ${\textbf{a}_2}=a(0,1)$ \cite{Sato,Mithun_2014}. In the following, we denote the spatial coordinates, time, condensate wave function, rotational frequency, and energies in units of $a_h$, $\omega_{\perp}^{-1}$, $a_h^{-3/2}$, $\omega_{\perp}$ and $\hbar\omega_{\perp}$, respectively, where $a_h=\sqrt{\hbar/m\omega_{\perp}}$. From Eq.(1) we obtain the 2D time-dependent coupled dimensionless GP equations (GPE) as 
\begin{eqnarray}
i{\partial \over \partial t}\psi_j(x,y,t)&=& \bigg[-{1\over 2}({\partial^2\over\partial x^2}+{\partial^2\over\partial y^2})+\tilde V_j(x,y)\nonumber \\
&+&\tilde g_{jj}|\psi_j(x,y,t|^2+\tilde g_{12}|\psi_{3-j}(x,y,t)|^2\nonumber \\
&-&\Omega L_z \bigg]\psi_j(x,y,t)
\end{eqnarray}
where $\tilde g_{jj}={4\pi Na_j\over a_h}\sqrt{{\lambda\over 2\pi}}$, $\tilde g_{12}={4\pi Na_{12}\over a_h}\sqrt{{\lambda\over 2\pi}}$, $\lambda ={\omega_z\over \omega_{\perp}}$ and $\tilde V_j(x,y)={1\over 2}(x^2+y^2) + V_{j_{,impurity}}(x,y)$. For random impurity potential $V_{1_{,impurity}}=V_{2_{,impurity}}=\sum_{n_1,n_2}V_{R}\exp\left\{{-[|\textbf{r} - {\textbf{r}_{{n_1},n_{2}}\mid^2}\over (\sigma/2)^2}\right\}$, where $V_{R}$ is drawn from the distribution $[-V_0,V_0)$. 

\section{Numerical details}
The split-step fast-Fourier method \citep{Mithun_2019} is used to solve the dimensionless coupled GPE equations (Eq.(2)) using imaginary time propagation. We consider the Thomas-Fermi wave function in the absence of rotation, $\psi_{\text{TF}}(x;\Omega=0)$, as the initial condition \cite{pethick2008bose}. Nevertheless, we adjust the wave functions of both components to be slightly different in order to make the initial condition asymmetric. We then introduce the rotation of desired frequency $\Omega$ to generate the vortices in the system. In the numerical simulations, we consider $512\times 512$ grid points for a domain size $32\times 32$ and $1024\times 1024$ grid points for a domain size $54\times 54$. We fix both the masses the same and $\tilde g_{11}=\tilde g_{22}=2000$  unless otherwise mentioned. Additionally, $N_1 = N_2$ is considered. Our motivating example is that of a mixture of 2D BECs of $^{87}$Rb atoms in the different
hyperfine spin states, the mass equality suggests our focus on a scenario of two hyperfine states of the
same gas, in particular $^{87}$Rb \cite{Mithun_2016}.  In the absence of rotation and optical lattice potential, the dimensionless chemical potential can be estimated from the expression $\tilde \mu=\sqrt{\frac{\tilde g_{11}+\tilde g_{12}}{\pi}}$. In full dimension, $\mu = \tilde \mu \hbar\omega_{\perp}$. The parameters varied  are the strength of the inter-component coupling $\delta={\tilde g_{12}\over \tilde g_{jj}}$, the rotational frequency $\Omega$ and the strength of the impurity potential. 
We calculate the structure factors profiles of the vortex lattices in terms of the spatial density of the condensate components as $S_j(\textbf{k})=\int dx dy |\psi_j(x,y,t)|^2e^{i\textbf{k} \cdot \textbf{r}}$, $j=1,2$. The structure factor profiles provide information about the periodicity of the condensate density. We also plot
the Delaunay triangulated lattice  of the condensate densities to determine the number of nearest neighbours of the vortex lattice to show lattice disorder through the creation of the lattice defects dislocations and disclinations. Dislocations  are lattice defects consisting of pairs of five-fold or seven-fold  and disclinations are  isolated five-fold or seven-fold coordinate axes respectively. For finding the dislocations and disclinations the boundary coordinates are not considered. In order to study the effect of  random pinning potential due to random impurities  or disorder  on the vortex lattice, 
we generate random potential $V_{impurity}$ by considering a square optical lattice $V_{lattice}(\textbf{r})$, where we fix the width of each Gaussian peak $\sigma$ to $0.5$ and distance between each peak, $a$ to 1. The height of each peak, $V_0$ has been changed with the help of random numbers which are uniformly distributed over $[-V_0, V_0]$  \cite{Mithun_2018}. 

\section{Effect of impurities or pinning centers on the equilibrium structures of unpinned vortex lattices}

\subsection{Effect of single impurity} 

 To see the effect of a single impurity on the vortex lattice structure, we consider the equilibrium vortex lattice in the presence of a single impurity. In the absence of any impurity, the rotating binary BECs with equal intra-component interaction but varying inter-component interaction and rotational frequency show rich equilibrium vortex lattice structures \citep{Kasamatsu}.

In the presence of an impurity, the vortex lattice structures are expected to get distorted due to the pinning of the lattice vortices with the impurity. We first show that the presence of even a single impurity can change the unpinned equilibrium vortex lattice structures of a binary BEC. We fix the impurity position $(x_0,y_0)$ near the trap center and in the middle of the two vortices of the unpinned triangular vortex lattice as it provides maximum distortion of the vortex lattice. Additionally, we set $V_{01}=V_{02}$.

\begin{figure}[h]
  \includegraphics[width=0.48\textwidth]{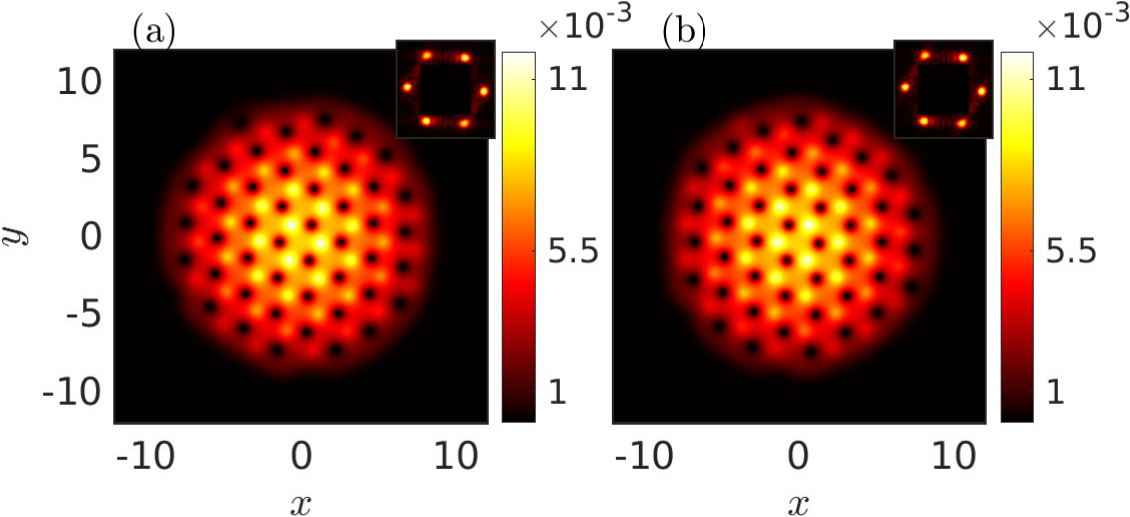}
 \caption{\label{Fig.1}{Condensate densities {with color bars} (a) $|\psi_{1}|^2$ and  (b) $|\psi_{2}|^2$ (right) without an impurity for $\delta=0.6$ and $\Omega= 0.71$.  The corresponding structure factor profiles are given in the inset.}}
\end{figure}

\begin{figure}[ht]
       \vspace{0pt}
 \includegraphics[width=0.48\textwidth]{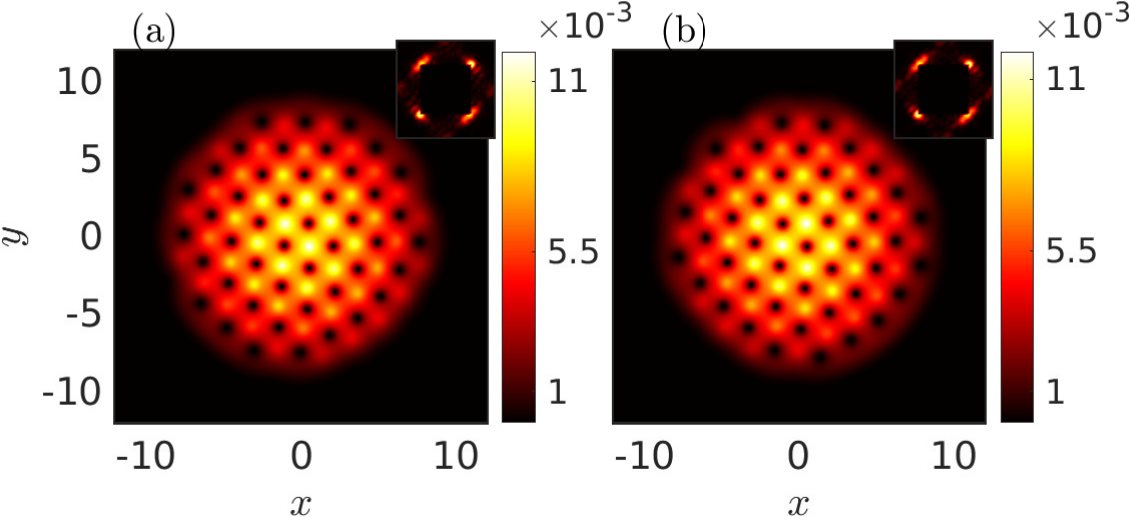}
     \vspace{0pt}
 \caption{\label{Fig.2}{ Condensate densities {with color bars} (a) $|\psi_{1}|^2$ and  (b) $|\psi_{2}|^2$ for $\delta=0.6$, $\Omega= 0.71$ in presence of a single impurity at $(x_0=0,y_0=-0.5)$ with strength $V_{01}=V_{02}=1$. The corresponding structure factor profiles are given in the inset.}}
\end{figure}

 We choose the interaction parameter between components $\delta$ and the rotational frequency $\Omega$, which corresponds to a triangular vortex lattice (\cite{Kasamatsu}).
 Fig.~\ref{Fig.1} shows the unpinned density profiles of the triangular vortex lattices of the two components. The corresponding structure factor profiles shown in the inset have six peaks as expected for a regular hexagonal Abrikosov lattice.
  
Interestingly, in the presence of a single impurity, the vortex lattice structures change from triangular to square lattice. This is shown in the density profiles in Fig.~\ref{Fig.2} and the corresponding structure factor profiles display four peaks as expected for a square lattice structure. 
This is because the vortices of both components near the impurity compete with each other to become pinned with the impurity. Since the impurity strength $V_0 \ll \mu$, neither of the vortices succeeds and, as a result, the entire vortex lattice rearranges to a square lattice to minimize the lattice potential energy $E_{lattice}=\langle \psi(x;\Omega)|V_{lattice}|\psi(x;\Omega)\rangle$.

\begin{figure}[ht]
       \vspace{0pt}
 \includegraphics[width=0.48\textwidth]{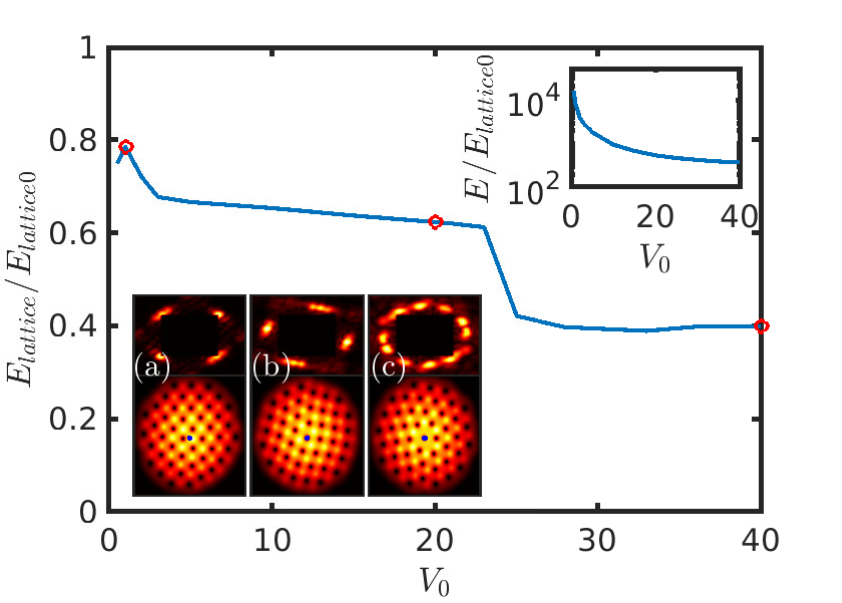}
     \vspace{0pt}
 \caption{\label{Fig.3}{Lattice energy versus impurity strength for a single impurity placed at $(x_0=0,y_0=-0.5)$. The corresponding vortex lattice and structure factors of one of the components corresponding to the red-marked points are given the bottom left inset (a-c), where the blue dot shows the position of a single impurity. The top right inset shows the normalized total energy $E/E{lattice0}$ calculated from the Eq.~\ref{eq:GPE_ener} as a function of the $V_0$. The other parameters are $\delta=0.6$ and $\Omega= 0.71$.}}
\end{figure}

To understand this transition, we calculated the equilibrium lattice structure starting from a vortex-free ground state and the corresponding lattice potential energies by varying the impurity strength. The results are shown in Fig.~\ref{Fig.3}, where $E_{lattice}$ is normalized with the $E_{lattice0}=\langle \psi(x;\Omega=0)|V_{lattice}|\psi(x;\Omega=0)\rangle$. As expected, in the presence of an impurity, the lattice potential energy decreases as a result of the pinning of the vortex with the impurity. As shown in the figure, there are three regimes with increasing strength of the pinning potential. In the first regime, we get higher $E_{lattice}$,  which remains nearly flat in the second regime. In the third regime, we see a lower $E_{lattice}$ for a further increase in impurity strength. In the first regime, for weaker strength of the impurity potential ($V_0 \ll \mu$) vortices of both components are weakly pinned, resulting in square lattices. In the second regime, $V_0$ from $ 3$ to $23$, we observe the pinning of the vortices of one of the components. However, the square-vortex lattice is unchanged. In the third regime, vortices of both components are pinned at impurity and as a result, we see maximum vortex lattice distortion. 

\begin{figure*}[ht]
       \vspace{0pt}
 \includegraphics[width=0.98\textwidth]{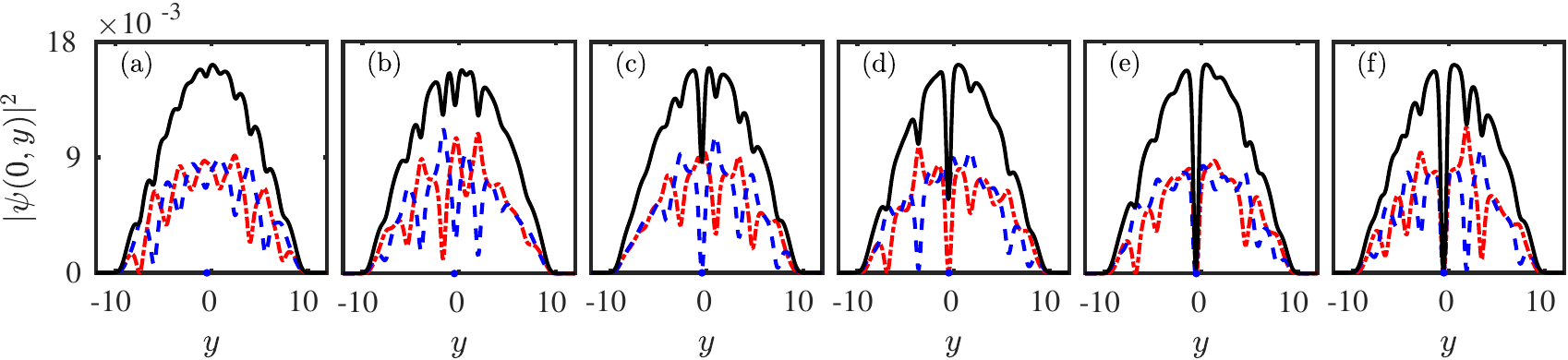}
     \vspace{0pt}
 \caption{\label{Fig.4}{ 1-Dimensional view of condensate densities $\rho_T(0,y)= |\psi_1|^2 + |\psi_2|^2$ (black), $|\psi_1(0,y)|^2$ (red)and $|\psi_2(0,y)|^2$ (blue) of the cases shown in Fig.~\ref{Fig.3} for $V_0=(1,2,10,20,28,40)$ (a-f). }}
\end{figure*}

Fig.~\ref{Fig.4} shows a cross-section ($x=0$ slice) of the density profiles as shown in Fig.~\ref{Fig.3}. The total density $\rho_T= |\psi_1|^2 + |\psi_2|^2$ is also shown in the same figure (black curve).   As mentioned above, for weak pinning the vortex lattice undergoes a transition to the square lattice (Fig.~\ref{Fig.2} and the first figure from the left in the density profiles in Fig.~\ref{Fig.3}) from the unpinned triangular lattice (Fig.~\ref{Fig.1}). A smoother total density $\rho_T$ is favourable for the square lattice \citep{Kasamatsu} which results in the shift of the positions of the vortex cores in such a manner that a peak in the density of one component is located in the density hole of the other, resulting in a decrease of $E_{lattice}/E_{lattice0}$. This is shown in (a) of the cross-section plots in Fig.~\ref{Fig.4}. With increasing strength of the pinning potential, the vortex lattice becomes more disordered resulting in fluctuations of the total density as shown in the Fig.~\ref{Fig.4}(b-f). It is to be noted that the decrease in $E_{lattice}/E_{lattice0}$ with well defined transition points is not quantified in the normalized total energy $E/E_{lattice0}$ as shown in the top right inset of the Fig.~\ref{Fig.3}. This further manifests that $E_{lattice}/E_{lattice0}$ is the right parameter to quantify a vortex matter transition in presence of a lattice potential \cite{Sato}.
 
\begin{figure}[ht]
       \vspace{0pt}
 \includegraphics[width=0.48\textwidth]{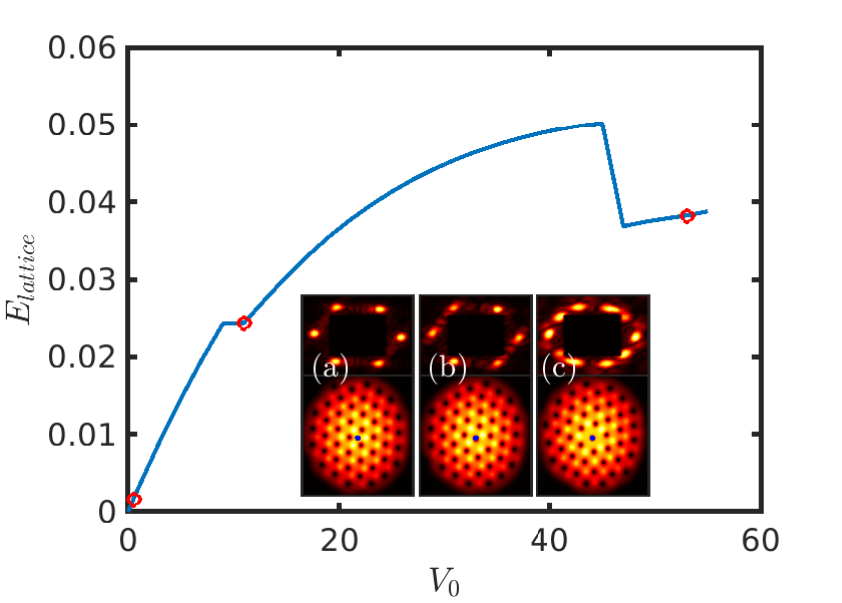}
     \vspace{0pt}
 \caption{\label{Fig.5}{ Lattice energy versus impurity strength for a single impurity placed at $(x_0=0,y_0=-0.5)$ starting from a vortex ground state shown in Fig.~\ref{Fig.1}. The corresponding vortex lattice and structure factors of one of the components corresponding to the red-marked points are given the inset (a-c), where the blue dot shows the position of a single impurity.}}
\end{figure}

An additional numerical experiment is carried out by considering the ground state of the vortex in Fig.~\ref{Fig.1} as the initial condition. The final vortex ground state of a $V_0$ is used as the initial condition for the simulation at $V_0+\epsilon$, where $\epsilon$ denotes a small increment in $V_0$. This is unlike the case shown in Fig.~\ref{Fig.3}, where the initial condition is always a vortex-free state $\psi_{\text{TF}}(x;\Omega=0)$.  The increase in impurity strength does not affect the geometry of the triangular lattice for $V_0 < 47$, as shown in the inset of Fig.~\ref{Fig.5}. This shows the presence of coexisting solutions for the same parameters set. Moreover, the order-disorder transition of the vortex lattice occurs at a higher strength compared to Fig.~\ref{Fig.3}.

\subsection{Effect of periodic impurities}
The vortex lattice structures of the rotating binary BECs can be controlled via the pinning of the vortices by periodic impurities or a periodic pinning potential. Experimentally such effects are created by loading the rotating  BECs on a co-rotating optical lattice \cite{Tung}. It is well-known that the addition of periodic artificial pinning centers helps to realize other vortex arrangements and also many dynamical phases \cite{laguna2001vortex,reichhardt2008moving}. In the context of superconductors, vortex pinning due to periodic pinning centers helps in increasing the critical current and controlling the fluxon dynamics \cite{aichner2019ultradense}. 

%
\begin{figure}[ht]
       \vspace{0pt}
 \includegraphics[width=0.48\textwidth]{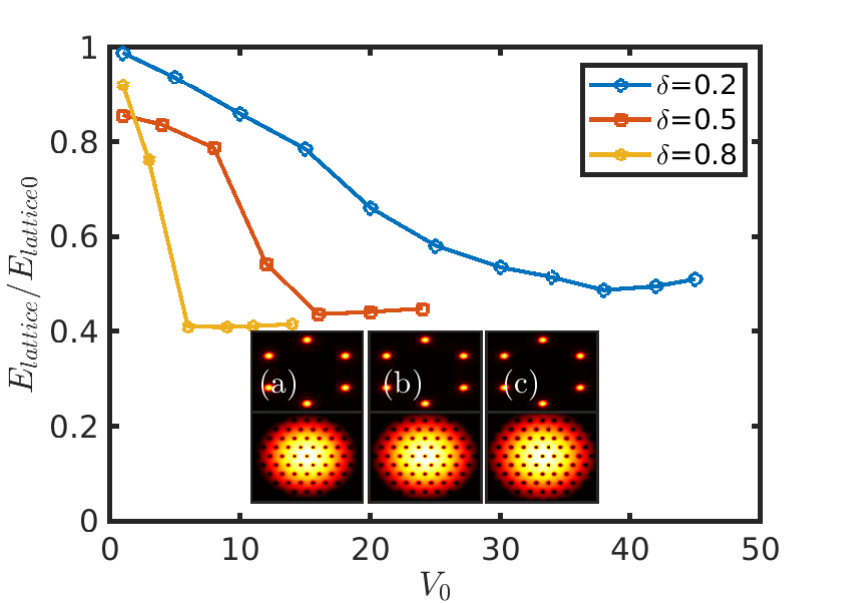}
     \vspace{0pt}
 \caption{\label{Fig.6}{Lattice energy versus impurity strength for a triangular optical lattice for $\delta=0.2$, $\delta=0.5$ and $\delta=0.8$ for the fixed rotation strength $\Omega=0.76$ and the lattice constant $a=2.28$. The condensate density of the first component $|\psi_1|^2$ in the inset shows the pinned triangular vortex lattices for (a) $\delta=0.2$, $V_0=11$, (b) $\delta=0.5$, $V_0=20$, and (c) $\delta=0.8$, $V_0=42$. The second component $|\psi_2|^2$ also exhibits the triangular vortex lattice.}}
\end{figure}

    \begin{figure}[htp]
 \begin{center}
      \centering    
      \includegraphics[width=0.48\textwidth]{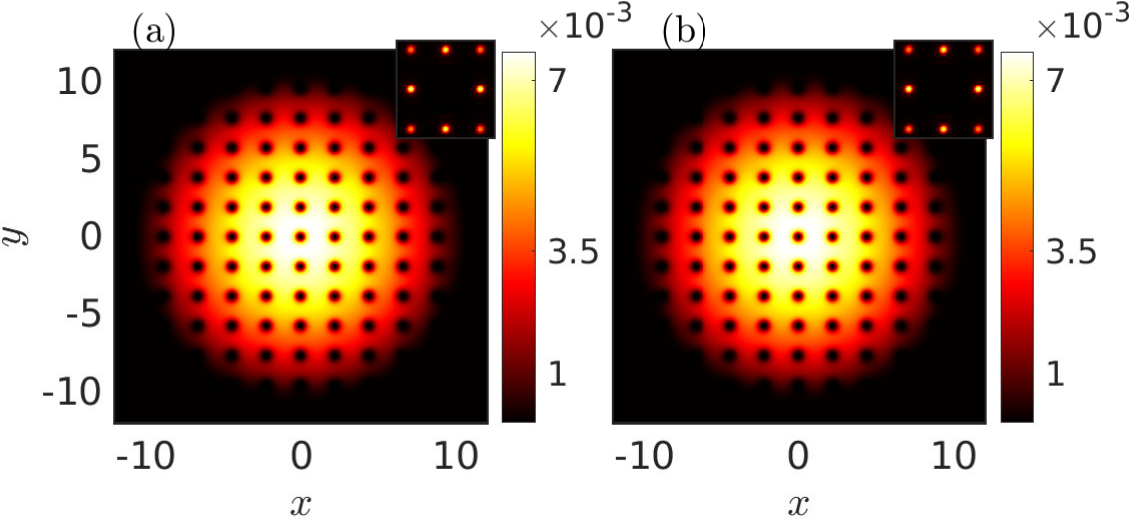}
 \caption{\label{Fig.8}{ The condensate densities {with color bars} (a) $|\psi_1|^2$ and (b) $|\psi_2|^2$ show the pinned square vortex lattices with lattice constant $a=2.2$ for $\delta=0.8$, $\Omega=0.76$ and $V_0=35$.} }
  \end{center}
  \end{figure}
  
We have simulated the coupled 2D Gross-Pitaevskii equations in the presence of optical lattices of triangular and square geometries. We show that the presence of weak periodic optical lattice potential leads to a transition from the unpinned vortex lattice structures to the structures of the optical lattice due to the pinning of the vortices by the optical lattice.
We show that for binary BECs the inter-component interaction $\delta$ plays a very crucial role in controlling the vortex lattice structures of each component. This is due to the difference in condensate densities produced by the inter-component interaction.


 In cases where the symmetries of the unpinned vortex lattice and the optical lattice are the same, we choose different lattice constants of the two lattices so as to demonstrate the perfect pinning of the unpinned vortex lattice to the optical lattice. After perfect pinning, the lattice constants of the original unpinned lattice match exactly with that of the optical lattice.

For example, for the choice of parameters $\delta=0.2$ and $\Omega=0.76$, the unpinned vortex lattice is triangular 
 (\cite{Kasamatsu}) with lattice constants determined by $a=2.2$ and therefore to show perfect pinning of the vortex lattice, we choose triangular optical lattice with different lattice constants for $a=2.28$. Similarly, for $\delta=0.7$ and $\Omega=0.76$ the unpinned vortex lattice is square (\cite{Kasamatsu})  with lattice constants determined by $a=2.12$ and accordingly to show pinning we choose square optical lattice with different lattice constants  $a=2.20$. 
We observe that a lower pinning strength of the optical lattice is required for pinning when there is a matching between the symmetries of the unpinned vortex lattice and the optical lattice. To show this we consider pinning of unpinned vortex lattice of various symmetries by a triangular optical lattice with lattice constants determined by $a=2.28$. For unpinned vortex lattice of different symmetries, we consider cases with increasing values of the inter-component interaction parameter $\delta$  as $0.2$, $0.5$, and $0.8$, keeping the rotational frequency the same as $\Omega=0.76$. For these cases, the corresponding unpinned vortex lattices are triangular, overlap, and square respectively (\cite{Kasamatsu}). The corresponding triangular pinned vortex lattices are shown in the inset of
Fig.~\ref{Fig.6}. From Fig.~\ref{Fig.6} we can see that increasing the strength of the triangular optical lattice is required for pinning as the symmetry of the unpinned vortex lattice changes from triangular to overlap to square. Fig.~\ref{Fig.6} further shows the respective lattice energies of the unpinned triangular and square lattices for various strengths of  the triangular optical lattice.  In both cases, the lattice energy decreases with the increasing strength of the optical lattice due to pinning. However, for the triangular lattice case, the lowering of the lattice energy is lesser as compared to that of the square lattice case due to the same symmetry of the unpinned vortex lattice and the pinning optical lattice.

 Fig.~\ref{Fig.8} shows the pinning of the unpinned square vortex lattice to the square optical lattice with lattice constants determined by $a=2.12$. For this, we take parameters values as $\delta=0.7$ and $\Omega=0.76$ which gives an unpinned square vortex lattice (\cite{Kasamatsu}) with lattice constants determined by $a=2.2$. Comparison of Fig.~\ref{Fig.6} and Fig.~\ref{Fig.8} shows that it requires more strength of the optical lattice to pin a square unpinned vortex lattice to a triangular optical lattice compared to the pinning of a square unpinned vortex lattice to a square optical lattice.
 
\section{Effect of random impurities or disorder}
\begin{figure}[htp]
 \begin{center}
  \includegraphics[width=0.48\textwidth]{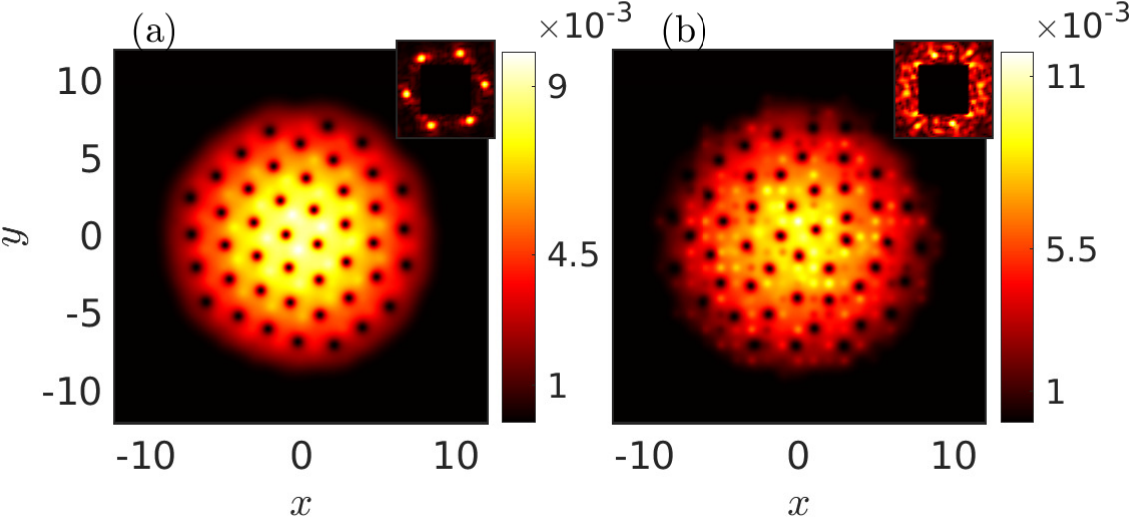}
 \includegraphics[width=0.48\textwidth]{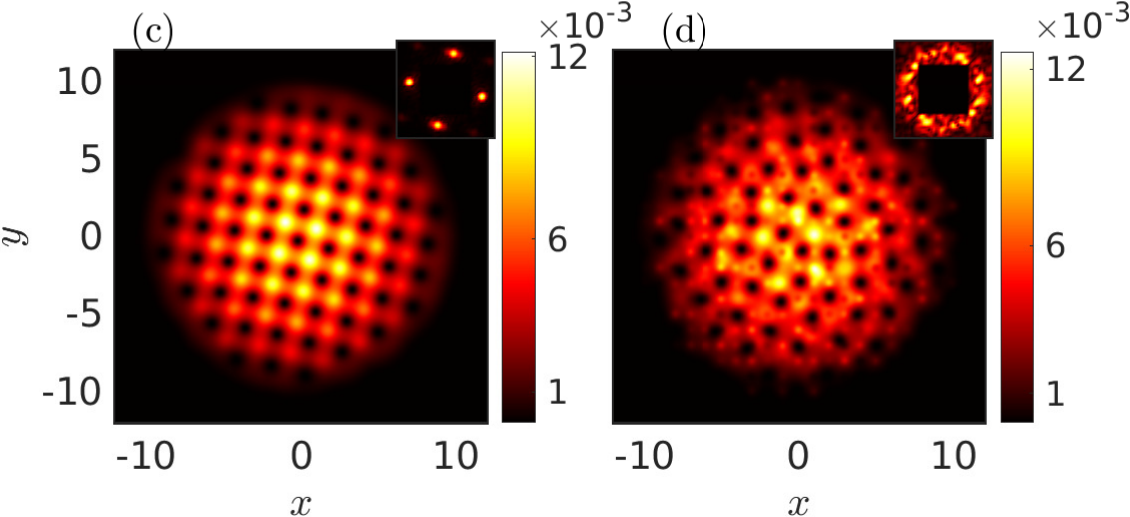}
 \caption{\label{Fig.9}{ Condensate density {with color bars} $|\psi_{1}|^2$ and structure factor (inset) for (a-b) $\delta=0.2$ and (c-d) $\delta=0.8$ for the fixed $\Omega$=0.76 with random impurity strength $V_{0j}=1$ for (a) and (c), $V_{0j}=10$ for (b) and $V_{0j}=15$ for (d) shows the order to disorder transition of a vortex lattice. The density $|\psi_{2}|^2$ shows similar pattern.}}
    \end{center}
  \end{figure}

In BECs the random impurities or disorder is created and controlled by the laser speckle method \cite{Ghosal}.
  For a single component BEC it has been shown that the vortex lattice melts with increasing strength of disorder due to pinning of the vortices with the random impurities \cite{Mithun_2016}.
  Also, for a fixed strength of the disorder and with increasing strength of the rotational frequency,
the vortex lattice gets increasingly disordered  leading to the melting of the vortex lattice. The  vortex lattice gets disordered by the creation of  lattice defects, dislocation, and disclination. Such melting of vortex lattice follows two steps. In the first step, the positional order of the unpinned vortex lattice disappears but the orientational order is retained. In the second step, both positional and orientational order disappears.  In the first step of melting, only dislocations are created whose number increases with increasing rotational frequency and the second step involves the creation of both dislocations and disclinations \cite{Mithun_2018}.

 The two-step vortex lattice melting is attributed to the Berezinski-Kosterlitz-Thouless-Halperin-Nelson-Young (BKTHNY) transition \cite{Kosterlitz_1973,Halperin,Nelson,Pauchard1996}. The two-step vortex lattice melting has been experimentally observed recently in type-II low-temperature superconductors \citep{Ganguli,Roy}. Recent numerical simulations of the dynamics of single component BEC in the presence of random impurities have also shown two-step vortex lattice melting \citep{Mithun_2018}. For the binary BEC, such studies of disorder-induced vortex lattice melting are important due to their relevance in the context of  the investigation of vortex lattice melting in more complex high-temperature superconductors having two order parameters.  Besides the possibility of melting the vortex lattice with increasing strength of disorder and rotational frequency similar to that of the single component BECs, we further show that for the binary BECs, it is also possible to melt the vortex lattices by varying the strength of the inter-component interaction $\delta$. 
 
   \begin{figure}[h]
    \begin{center}
       \includegraphics[width=0.48\textwidth]{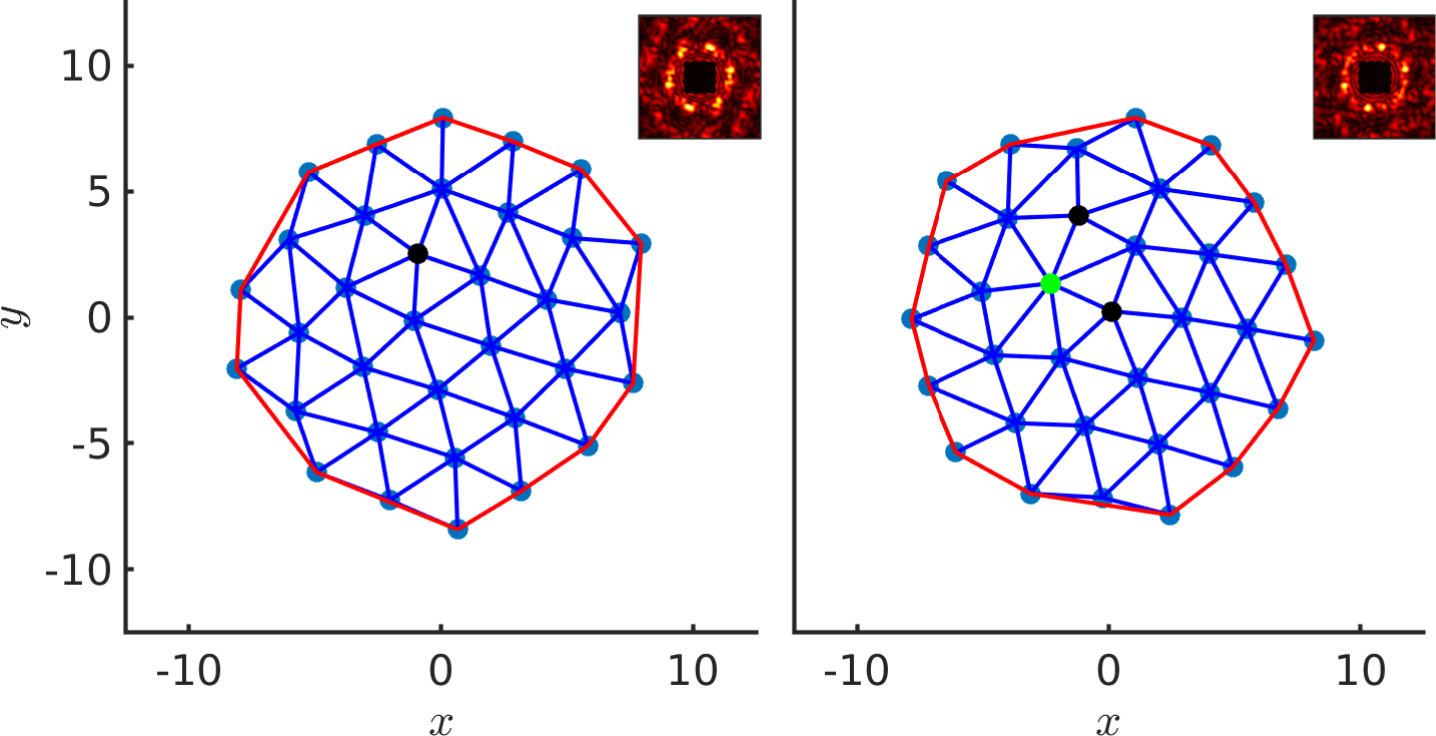}  
          \includegraphics[width=0.48\textwidth]{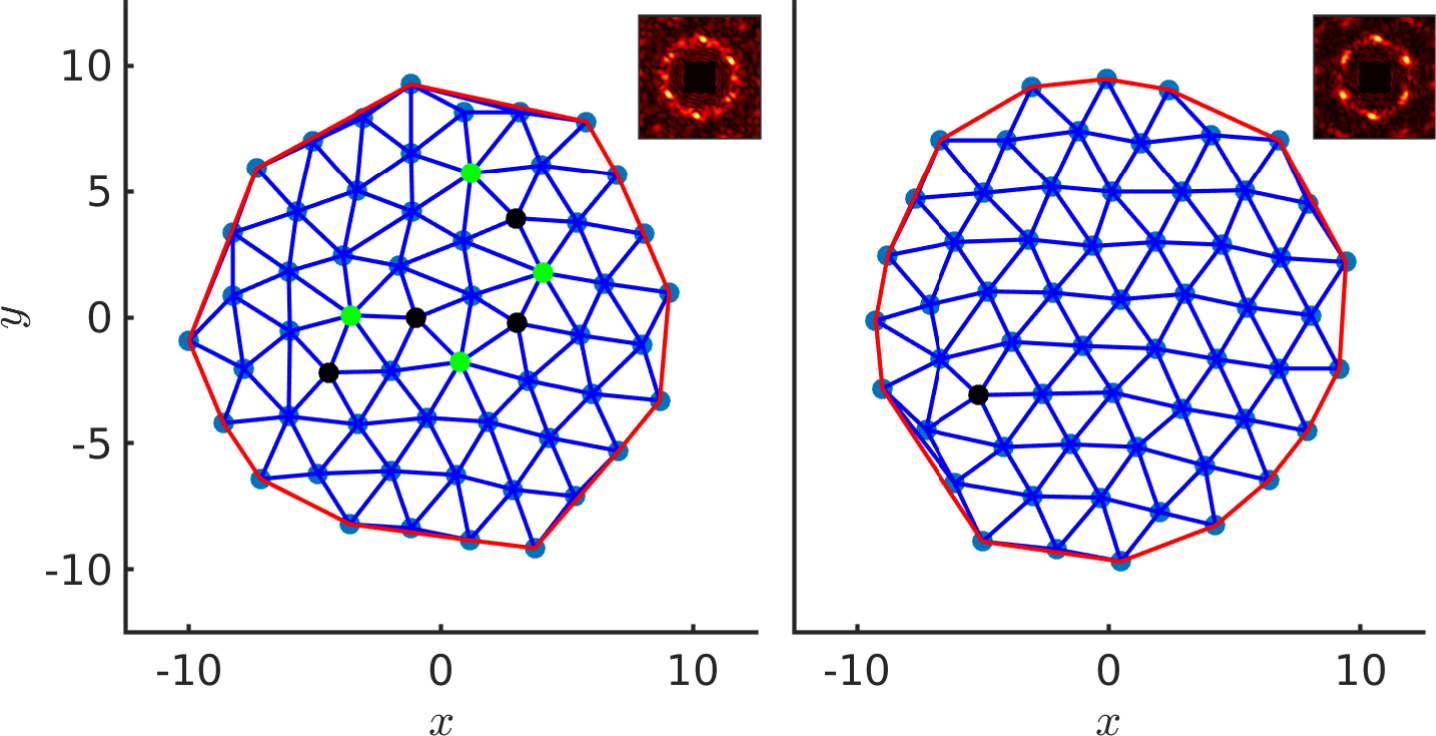}
     \includegraphics[width=0.48\textwidth]{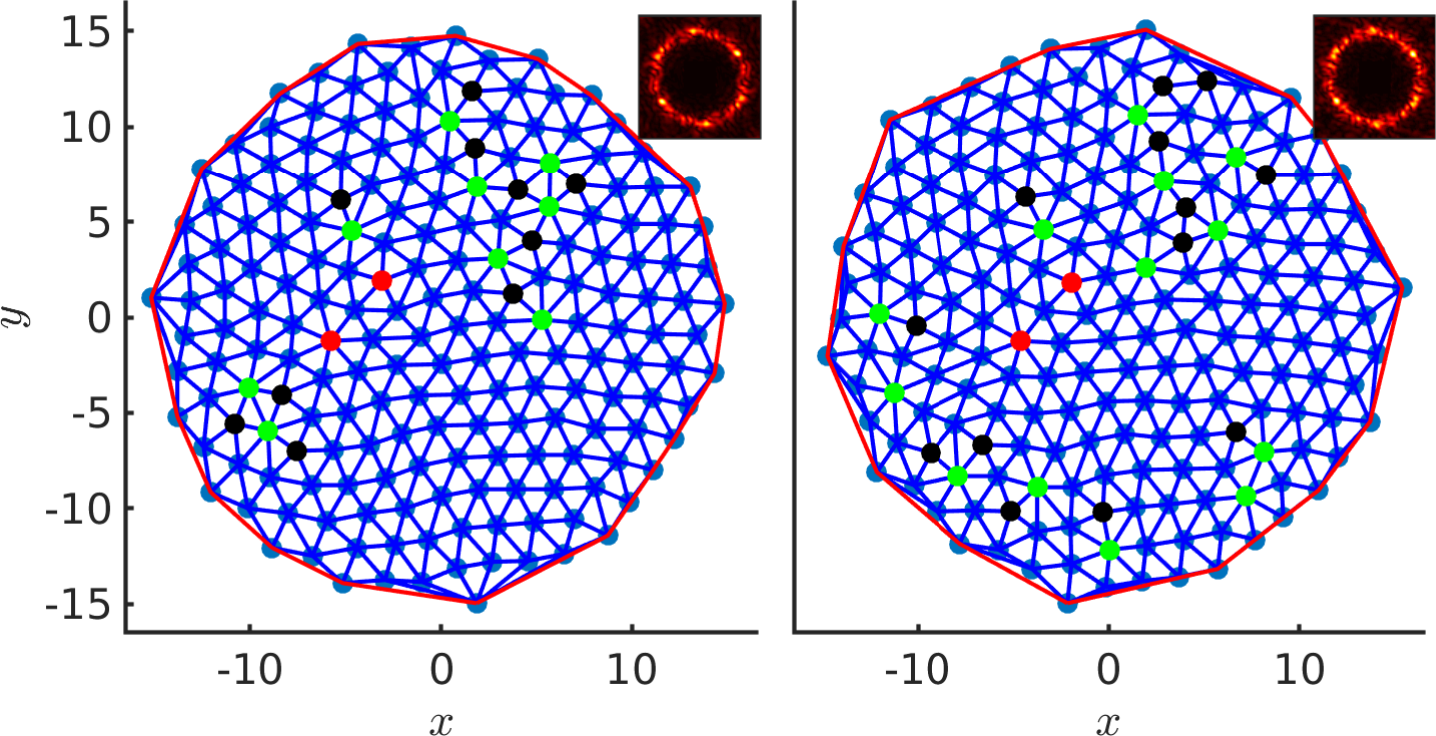}
\caption{Top, middle and bottom panels are Delaunay triangulated disordered vortex lattice of the first (left) and second (right) BEC components for $\delta=0.2$, $g_{11}=8000$ and $\Omega =( 0.45,0.6,0.9)$ respectively and random impurity strength $V_{0j}=1.5$. }
   \label{Fig.10}
\end{center}
  \end{figure}

  \begin{figure}[!htbp]
         \includegraphics[width=0.48\textwidth]{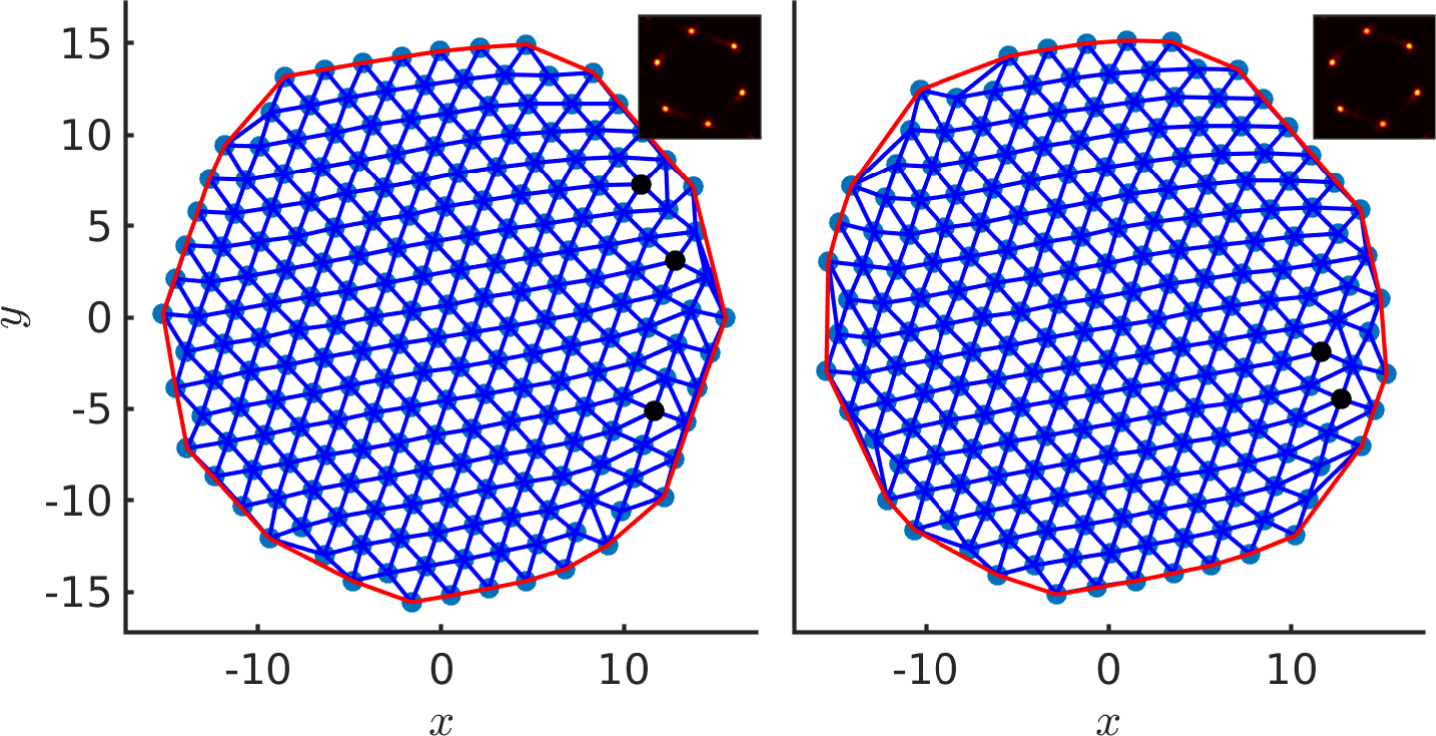}  
          \includegraphics[width=0.48\textwidth]{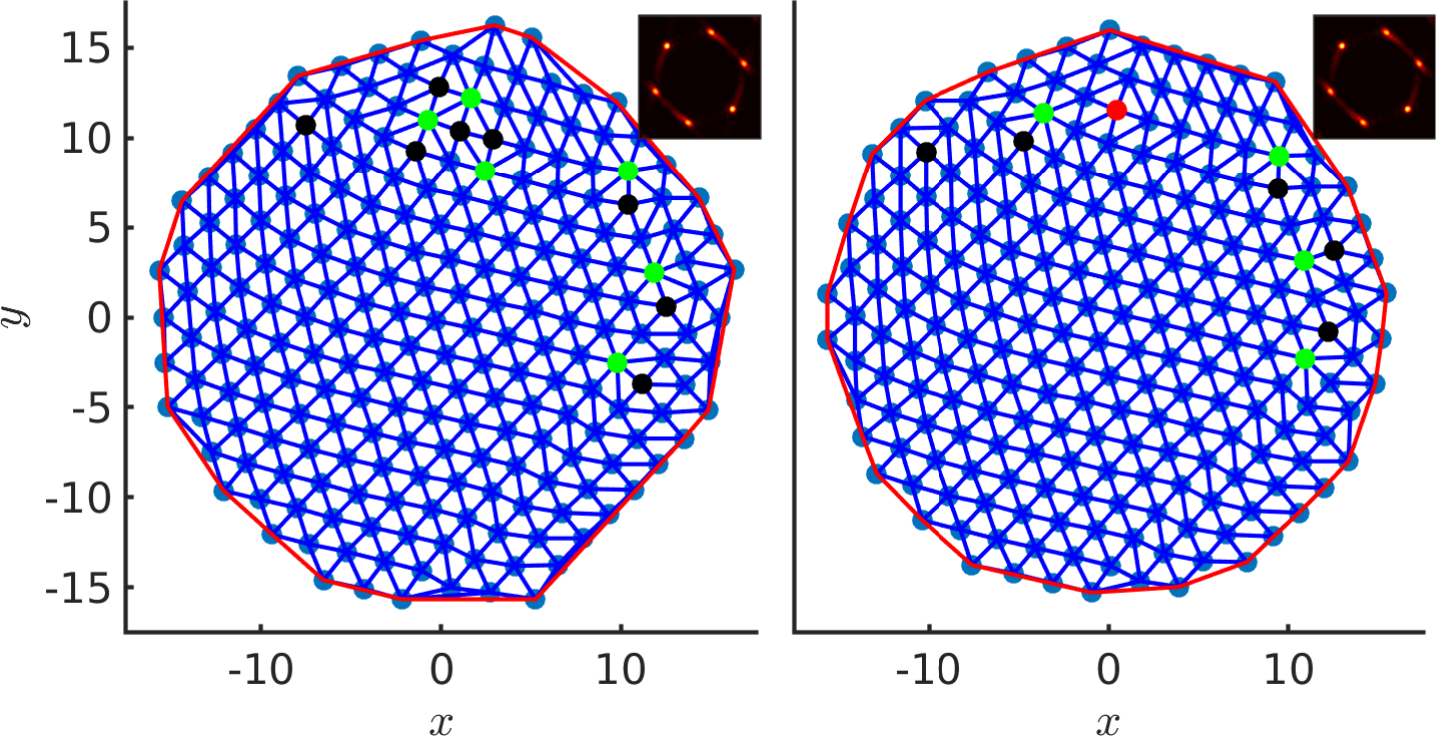}
     \includegraphics[width=0.48\textwidth]{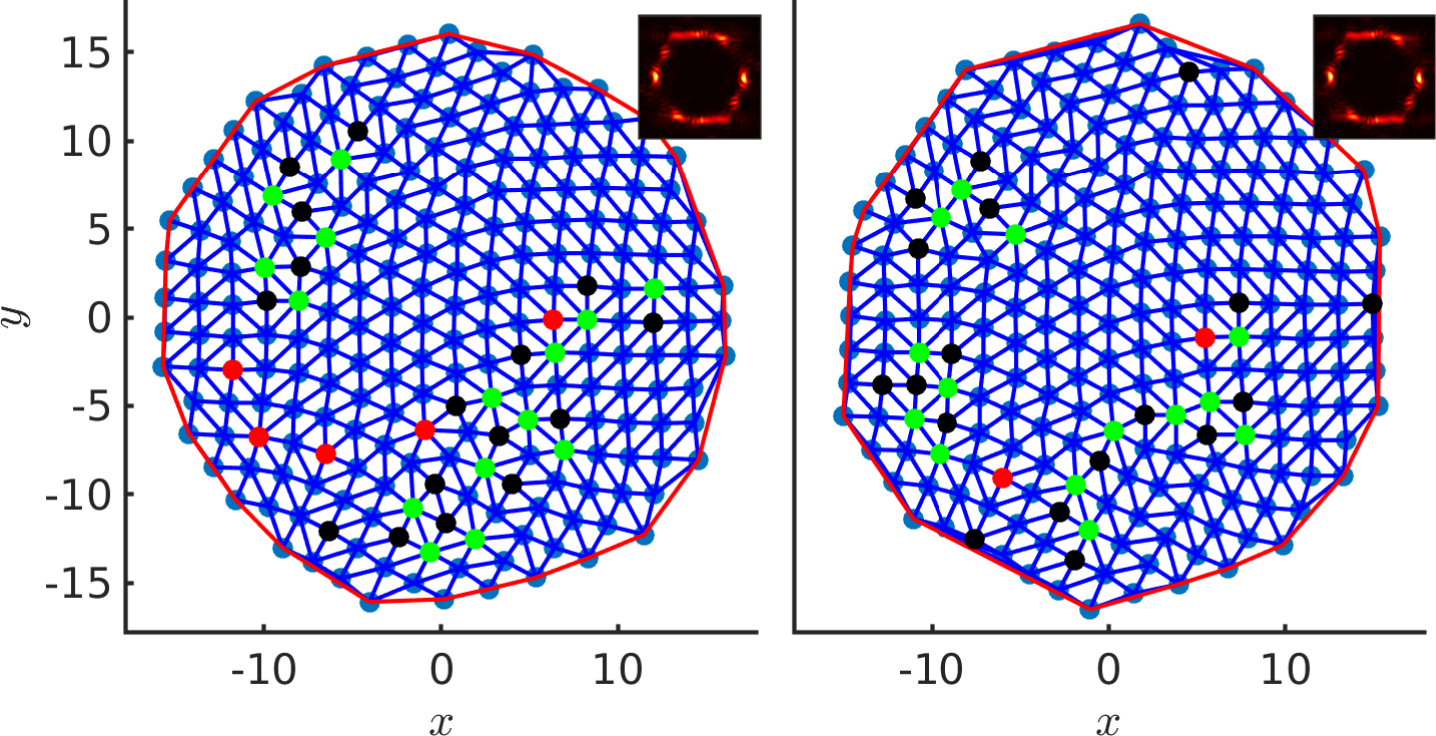}
\caption{Top, middle and bottom panels are delanuay triangulated disordered vortex lattices of the first (left) and second (right) BEC components for $\Omega=0.9$, $g_{11}=8000$ and $V_{0j}=0.5$ with $\delta = (0.45,0.55,0.67)$ respectively.}
\label{Fig.11}
  \end{figure} 

To show this we consider random impurities effects on both the triangular as well as the square vortex lattice regimes.
Fig.~\ref{Fig.9} shows that for the disorder strength $V_{01}= V_{02}=1$, the vortex lattices of both the components remain nearly hexagonal and square respectively for the cases $
\delta=0.2$ and 0.8 as seen from the density plots and the corresponding structure factor profiles which shows six periodic peaks.
But at higher disorder strength, the vortex lattice of both the components
gets  disordered as seen from the  density plots and the structure factor profiles.

Similar to the single component BECs, the two-step vortex lattice melting  by the creation of lattice defects with increasing rotational frequency \cite{Mithun_2018} is also observed for the binary BECs. 
The corresponding Delaunay triangulated disordered vortex lattice structures are shown in Fig.~\ref{Fig.10}. In the plots, the lattice defects dislocations with five-fold and seven-fold nearest coordinates are shown in black and green-filled circles respectively and the disclinations are shown in red-filled circles. 
From Fig.~\ref{Fig.10} we can see that for $\Omega=0.45$  and $\Omega=0.6$ the  lattice defects present for both the components are only the dislocations. In this case, the translational symmetry of the unpinned triangular lattice is lost but the rotational symmetry is still present. This can be seen from the structure factor profiles in the inset of the figures of the corresponding densities for these two cases. 
The structure factor profiles for both components show six nearly periodic  peaks implying that the rotational invariance of the unpinned triangular lattice is still maintained even in the presence of the random impurities.  
As the rotational frequency increases further to $\Omega=0.9$, both types of lattice defects, dislocations, and disclinations are present. The number of lattice defects increases  leading to the melting of vortex lattices for both components. With the appearance of disclinations, the rotational invariance is also lost. The structure factor profiles in the inset show that there are more intense peaks corresponding to the disordered vortex lattices. 

Similarly, the two-step vortex lattice melting is also observed for different random realizations and the corresponding plots are not shown here to avoid cluttering of the figures. 
\begin{figure}[ht]
       \vspace{0pt}
 \includegraphics[width=0.48\textwidth]{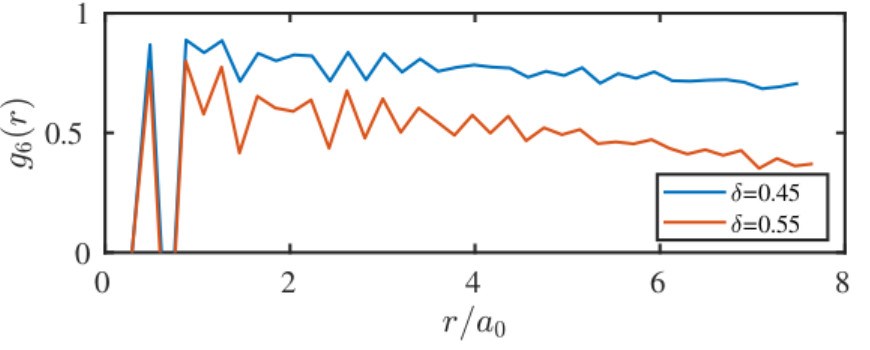}
     \vspace{-15pt}
 \caption{\label{Fig.12}{The orientational correlation function $g_6(r)$ as a function of $r/a_0$ for $\delta=0.45$ and  $\delta=0.55$ for a fixed rotation strength corresponds to the Fig.~\ref{Fig.11}.}}
\end{figure}

We have also observed the two-step vortex lattice melting for the binary BECs with increasing strength of the inter-component interaction $\delta$. This is shown in Fig.~\ref{Fig.11}. From the structure factor profiles in the inset of the figure, we can see that for the inter-component interaction strengths $\delta = 0.45$ and $\delta = 0.55$ the six-fold rotational symmetry is still preserved.  The loss of long-range order with increase in $\delta$ can be verified from the orientational correlation function $g_6(r) = \left (\sum_{i,j}\Theta \left (\frac{\Delta r}{2} - \left |r - \left |{\mathbf r}_i - {\mathbf r}_j \right |\right |\right ) \cos 6 \left (\theta ({\mathbf r}_i) - \theta ({\mathbf r}_j)\right )\right )\times(1/n(r,\Delta r))$, where $\Theta (r)$ is the Heaviside step function, $\theta ({\mathbf r}_i) - \theta ({\mathbf r}_j)$ is the angle between the bonds located at ${\mathbf r}_i$ and the bond located at ${\mathbf r}_j$, $n(r,\Delta r) = \sum_{i,j}\Theta \left (\frac{\Delta r}{2} - \left |r -\left |{\mathbf r}_i - {\mathbf r}_j\right |\right |\right )$, $\Delta r$ defines a small window of the size of the pixel around $r$ and the sum is over all the bonds, shown in Fig.~\ref{Fig.12}.  The range of variable $r$ is determined by the lateral size of each image. In Fig.~\ref{Fig.12}, we restrict the $r$ to half the lateral size of the image which corresponds to approximately 7.5$a_0$ (where $a_0$ is the average lattice constant) for $\delta = 0.45$ and 7.6$a_0$ for $\delta = 0.55$. The sharp peaks in Fig.~\ref{Fig.12} corresponds to the nearest neighbour bond distances. Though a power-law decay of the orientational order is expected as a characteristic of a quasi-long-range orientational order \cite{chandra2015disordering}, the small range of radial distance is not providing any conclusive evidence. The small decay rate of $\delta = 0.45$ as compared to the case of $\delta = 0.55$ indicate that better orientational order of $\delta = 0.45$ case.  When the inter-component interaction strength is increased further to $\delta = 0.67$, both translational and rotational symmetries are lost. The lattice structures of both components become completely disordered, as shown in the corresponding structure factor plots depicted in Fig.~\ref{Fig.11}. We have verified similar results for different random realizations. 

\section{Summary, Conclusions and Future challenges}
We have studied how the vortex lattice structures of binary BECs can be controlled by the pinning of the vortices by impurities or disorder. We have considered the pinning effects of three different types of impurities, all of which can be created experimentally using laser beams. By numerically solving the time-dependent coupled Gross-Pitaevskii equations we have observed the transition of the ordered unpinned vortex lattice structures to various phases where vortices order in lattice structures of different symmetries, in  periodic arrays, as well as in completely disordered or melted lattice. The transitions are determined by the competition between the strengths of the inter-component interaction, the impurity potential, and the rotational frequency. The intercomponent interaction plays a very important role in inducing different degrees of disorder in the vortex lattices of the two components due to differences in the densities of the condensate. 

In the presence of a single impurity, we observe the transition of the unpinned vortex lattice structure from the triangular to square and also to a disordered lattice. For periodic impurities, we have considered the triangular and square optical lattices co-rotating with the binary BEC. We observed the transition of the unpinned vortex lattices to the pinned lattices where all the vortices are pinned to lattice points. The minimum pinning strength of the optical lattice is required to pin the vortex lattice if there is a matching between the symmetries of the unpinned vortex lattice and the optical lattice. However, it requires a lesser strength of the optical lattice potential to pin a triangular vortex lattice to a triangular optical lattice as compared to the pinning of a square vortex lattice to a square optical lattice. Also, it requires higher strength of the optical lattice potential to pin a triangular vortex lattice to a square optical lattice as compared to the pinning of a square unpinned vortex lattice to a square optical lattice. In the presence of random impurities or disorder, the unpinned vortex lattice melts. The melting and loss of long-range order occur following a two-step melting process by the creation of an increasing number of lattice defects with increasing rotational frequency as well as the strength of the random pinning potential. Interestingly, we observed that similar vortex lattice melting can also occur in  binary BECs by increasing the strength of the inter-component interaction for a much weaker strength of the random pinning potential. 

In conclusion, rotating binary BECs in the presence of impurities or disorder provide an interesting system for studying new quantum phases of matter as well as phenomena known from condensed matter in new perspectives. 
In this context, the results of the impurity-induced vortex lattice structures in binary BECs as reported here are relevant for vortex dynamics in impurity-doped high-temperature superconductors having two order parameters and should be observed when the experimental results on the vortex lattice structures in such complex superconductors are available in the future.

An interesting direction for future work would be a detailed study of the pinned phases of the binary BECs in the presence of recently realized quasicrystalline optical lattices \cite{Viebahn}. In comparison to their periodic counterparts, the aperiodic nature of the underlying quasiperiodic potential and the intercomponent interaction are expected to create various intriguing vortex lattice phases. Another possible direction of future research would be to study the dynamical phases of the vortices in periodic pinning potentials that are distinct from the triangular and square potentials. Experiments with periodic pinning arrays such as honeycomb and Kagome revealed interesting anomalies in critical currents \citep{Morgan}.  Numerical simulation of vortices in honeycomb optical lattice has shown a remarkable variety of dynamical phases that are distinct from triangular and square pinning arrays and can flow in a direction of the driving force due to the depinning of vortices leading to transport \cite{Reichhardt}. Similar studies for binary BECs will be relevant in the context of experimental observation of anomalous critical current behaviour or the peak effect in high-temperature superconductors. 
\section{Acknowledgments}
The authors would like to thank Panayotis G. Kevrekidis for fruitful discussions. BD would like to thank the Science and Engineering Research Board of India for funding the research, Grant No. CRG/2020/003787.  
\bibliographystyle{apsrev4}
\let\itshape\upshape
\bibliography{bec_main}
\end{document}